\documentclass[12pt]{article}
\usepackage{calrsfs,bbm,bm}
\usepackage{axodraw,color,dsfont,cancel,graphicx,subfig}
\usepackage{amsmath}
\usepackage{amssymb}

\def\centerbox#1#2{\centerline{\epsfxsize=#1\textwidth\epsfbox{#2}}}

\usepackage{exscale}

\font\fourteenbf=cmbx12 scaled\magstep1

\def\a0size{6}

\newcommand{\lsi}{\raise0.3ex\hbox{$<$\kern-0.75em\raise-1.1ex\hbox{$\sim$}}}
\newcommand{\gsi}{\raise0.3ex\hbox{$>$\kern-0.75em\raise-1.1ex\hbox{$\sim$}}}
\newcommand{\lsim}{\mathop{\lsi}}

\renewcommand{\vec}[1]{{\bm #1}}

\newcommand{\be}{\begin{equation}}
\newcommand{\ee}{\end{equation}}
\newcommand{\baed}{\begin{aligned}}
\newcommand{\eaed}{\end{aligned}}

\begin{document}
 
\setlength{\baselineskip}{0.6cm}
\newcommand{\figysize}{16.0cm}
\newcommand{\figtopspace}{\vspace*{-1.5cm}}
\newcommand{\figbottomspace}{\vspace*{-5.0cm}}
  

\begin{titlepage}
\begin{flushright}
BI-TP 2010/48
\\
December 2010
\\
\end{flushright}
\begin{centering}
\vfill

{\fourteenbf 
\centerline{ 
Thermal production of relativistic Majorana neutrinos: 
      }
\centerline{ 
Strong enhancement 
by multiple soft scattering
      }
}

\vspace{1cm}

Alexey Anisimov, 
Denis Besak \footnote{dbesak@physik.uni-bielefeld.de}, 
Dietrich B\"odeker  \footnote{bodeker@physik.uni-bielefeld.de}

\vspace{.6cm} { \em 
Fakult\"at f\"ur Physik, Universit\"at Bielefeld, D-33615 Bielefeld, Germany
}

\vspace{2cm}
 
{\bf Abstract}

\end{centering}
 
\vspace{0.5cm}
\noindent
The production rate of heavy Majorana neutrinos is relevant for models of
thermal leptogenesis in the early Universe.
In the high temperature limit
the production can proceed via the $ 1\leftrightarrow 2 $ (inverse) decays
 which are allowed by the
thermal masses. 
We consider new production mechanisms which are obtained by 
including additional soft gauge interactions with the plasma. 
We show that an arbitrary number of such interactions gives 
leading order contributions, and we sum all of them.
The rate turns out to be smooth in 
the region where the $ 1 \leftrightarrow 2 $  
processes 
are kinematically forbidden. At higher temperature
it is  enhanced by a factor 3 compared to 
the $ 1 \leftrightarrow 2 $ rate.

\vspace{0.5cm}\noindent

 
\vspace{0.3cm}\noindent
 
\vfill \vfill
\noindent
 
\end{titlepage}
 
\section{Introduction and motivation}
\label{sc:introduction} 

One of the outstanding problems of standard cosmology is to explain the origin
of the asymmetry between matter and antimatter. Without such an asymmetry, all
the structures we observe today would have never formed and mankind would not
exist. The asymmetry can be expressed as the \emph{baryon-to-photon ratio} 

\be
\frac{n_B}{n_\gamma} = (6.21 \pm 0.16 )\cdot 10^{-10}
\ee
whose numerical value is obtained from a combined analysis of data for
large-scale structure and the spectrum of the Cosmic Microwave Background~\cite{WMAP}. 

In order to obtain a net baryon asymmetry,
only three conditions need to be met, as outlined by Sakharov in his
seminal paper~\cite{Sakharov}. Yet, providing a model that can 
successfully explain the measured
baryon-to-photon ratio remains a challenging task.
Several different scenarios how to realize the
Sakharov conditions have been devised~\cite{Baryogenesis}. 
In the last decade, \emph{leptogenesis}~\cite{Fukugita}
has become very popular. The basic idea of most leptogenesis models is to enlarge the particle content of
the Standard Model (SM) with \emph{heavy Majorana neutrinos}. In the simplest
realization they interact with the SM particles via a Yukawa coupling to ordinary,
left-handed leptons and the Higgs bosons as follows: 
\begin{align}\label{Lint}
   \mathcal{L}_{\rm int} = h_{ij} \overline{N}_i {\widetilde \varphi}^\dagger 
   \ell_{Lj}
   + \mbox{ h.c. },
\end{align}
where $N$ stands for the Majorana neutrinos, $\ell_L$ and $\varphi$ are the left-handed
lepton doublet and the Higgs doublet, and $\widetilde \varphi \equiv
\varepsilon \varphi ^ \ast $ with $ \varepsilon _ { \alpha \beta } = -
\varepsilon _ { \beta \alpha }$, $ \varepsilon _ { 12 } = 1 $.
Finally,  the indices $i,j$ label
the fermion families, and $h_{ij}$ is the Yukawa coupling matrix which need
not be diagonal.

The Majorana neutrinos are unstable and decay both into leptons and
antileptons, $N \to \ell  \varphi $, 
$ N \to \bar {\ell}  \varphi^\dag$. The CP symmetry is violated and
the corresponding decay rates are not equal. Therefore, an   
excess of antileptons over leptons can be generated. The resulting asymmetry
is converted into an excess of baryons over antibaryons via the sphaleron
transitions 
which  conserve $B -
L$ but violate $B + L$~\cite{Shaposhnikov}. In addition to providing a source
for the measured baryon asymmetry, this scenario offers a framework to
explain the smallness of the neutrino masses via the \emph{seesaw
  mechanism}~\cite{seesaw}. 
This twofold virtue is what makes the scenario of leptogenesis
particularly appealing. 

Despite a substantial amount of work and
progress~\cite{CP,Giudice,Kiessig,CTP} 
a complete theory of leptogenesis is still lacking \cite{anisimov}.
In this paper we study a type of processes which so far has not been
considered in this context. We show that they contribute to the
production of heavy Majorana neutrinos at leading order in the
coupling constants, and we find that their contribution is numerically
large.
Our results therefore constitute an important step towards
a complete treatment of thermal leptogenesis.

We compute the production rate of a Majorana neutrino in a hot electroweak
plasma that is fully equilibrated, except for 
the Majorana neutrinos themselves. The production rate is part of the network of
Boltzmann equations that is solved to obtain the baryon asymmetry.
Since it sets the initial conditions for leptogenesis, it is also
of practical interest to study the production rate by itself. 
We assume that the number density of Majorana neutrinos is
small compared to the equilibrium density, so that the inverse processes which 
reduce their number density can be neglected. We focus on the
lightest Majorana neutrinos $N_1 \equiv N$, which we assume to be the
dominant source of lepton asymmetry. 
 
When the temperature $ T $ is sufficiently 
above the Majorana neutrino mass $ M _ N $,
a peculiar type of production mechanism occurs. It was already considered in 
Refs.~\cite{Giudice,Kiessig}. Interactions with the hot plasma generate thermal
masses, which are much bigger for SM particles 
than for the Majorana neutrinos.
Therefore the SM particles can become
``heavier'' than the Majorana neutrino, and the decay of a Higgs boson into
Majorana neutrino and SM lepton can become possible.
 Since thermal masses are parametrically small
compared to the typical particle momentum, all momenta involved in
this decay are nearly collinear. In this paper we show that there are
additional nearly collinear processes, involving soft electroweak gauge
interactions, which contribute to the leading
order production rate.

We focus on the leading order in the SU(2) and U(1) gauge couplings $g$ and
$g'$, the top
quark Yukawa coupling constant $h _ t$ and the Higgs self-coupling
$\lambda$. We do not consider the production via $ 2 
\leftrightarrow 2 $ scattering processes \footnote{In the literature 
one can find  several calculations of 
$ 2 \leftrightarrow 2 $ scattering rates (see, e.g.
\cite{Giudice,pilaftsis,pedestrians,davidson,hahn-woernle}). 
However, to the best of our 
  knowledge, there is no calculation which consistently treats all leading
  order thermal effects.}.
For the power counting we assume that all these couplings
are of the same order and collectively refer to them by $g$. 
All other SM couplings are neglected.
 We perform the computation in the high-temperature
regime where $M_N \ll T$. This allows us to formally treat the mass of
the Majorana neutrino as being soft, $ M_N \sim g T $, 
 and therefore  parametrically of the same order as 
the thermal Higgs and lepton masses. We demonstrate 
that even at leading order
the production cannot simply be understood in terms of scattering
processes involving only a handful of particles.  '$ N
$-strahlung' and inverse decay  processes involving multiple
interactions mediated by  soft electroweak 
gauge bosons are not suppressed despite the
large number of vertices.
The emission occurs almost collinearly, so that propagators are
nearly on-shell and compensate the suppression. In position space the
radiated particle and its source overlap over large distances, and the
interference of different interactions cannot be
neglected~\cite{landaumigdal}. This phenomenon has been studied in various
contexts such as parton energy loss \cite{baier}, 
photon~\cite{AurencheLPM,arnoldPhoton,BesakBodeker}
 and gluon~\cite{arnoldGluon} production in a quark-gluon
plasma (for a general discussion see~\cite{arnoldKinetic}). 
Recently~\cite{BesakBodeker}, we presented a new approach how to
consistently include soft gauge interactions in the computation of a
thermal particle production rate. It is formulated in a way that is
largely independent of the type of particles whose production we want
to study, and can therefore easily be adapted to the case at hand. 

The paper is organized as follows. In Sec.~\ref{sc:rate} we relate the
production rate, which describes out-of-equilibrium physics, to a real-time 
correlation function in equilibrium. The latter can be calculated
in thermal field theory, which is subject to the following sections.
We describe the physics of collinear emission and outline the relevant
momentum scales in Sects.~\ref{sc:processes} and \ref{sc:kinematics}.
In Sec.~\ref{sc:strategy} we give a short and qualitative summary how
we proceed to obtain the leading order production rate due to collinear emission processes.
The rest of Sec.~\ref{sc:LPM} provides all the technical details that are needed
to arrive at the final results, and the reader who is not interested
in the details of their derivation may skip directly to 
Eq.~\eqref{ratepsif}. In Sec.~\ref{sc:Numerical} we present numerical
results and we conclude in Sec.~\ref{sc:conclusions}. The appendix
finally explains how to obtain the numerical solutions.

\section{Production rate 
and thermal field theory}      
\label{sc:rate} 

At lowest order in their Yukawa couplings  
the production
rate of Majorana neutrinos with 4-momentum $k$ can be written as
 \footnote{Our formula for the production rate of a spin 1/2-fermion is
  consistent with those shown in the literature (e.g.~\cite{BolzPradler})
  although the overall sign appears to be different. However, with our
  conventions the self-energy corresponds to $(-1)$ times the Feynman
  diagrams.} 
\begin{align}
   \label{rate} 
   \frac{\text{d}\Gamma}{\text{d}^3 k} =
   \frac{1}{(2\pi)^3 k^0} f _ { \rm F } (k^0) \operatorname{Tr} \left[
     \cancel{k}  \operatorname {Im} \Sigma_{\rm ret} (k) \right].
\end{align}
Here $\Sigma_{\rm ret} (k) = \Sigma ( k ^ 0 + i 0^+, \vec k ) $ is
the retarded self-energy for the Majorana neutrinos, 
and 
$ f _ { \rm F } $ is the Fermi-Dirac distribution. 

\begin{figure}[t] 
  \centering
\begin{picture}(50,90)(130,20)

\SetWidth{1}

\Vertex(180. ,60. ){2}
\Line(180,60)(210,60)

\ArrowArc(150,60)(30,0,180)
\DashArrowArc(150,60)(30,180,0){5}

\Vertex(120. ,60. ){2}
\Text(80,60)[]{ $k$}
\Line(120. , 60.)( 90., 60.)

\Text(150,105)[]{ $p$}

\end{picture}

\caption{The imaginary part of this diagram gives the production rate
due to decay and inverse decay processes. 
This diagram also 
determines the inhomogeneous term in Eq.~(\ref{72}).}
\label{fg:2point}
\end{figure}
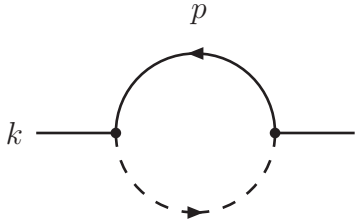
Due to the Majorana-nature of $ N $ there are two types of diagrams
contributing to the self-energy which differ by the orientation of the
internal fermion lines. We neglect Standard Model CP-violation. Then
both types of diagrams give the same contribution.  Therefore we consider only
one of them, and multiply by 2 to obtain the correct rate. We
do this by considering only the self-energy for right-handed $ N $'s,
so that the non-vanishing components of the self-energy fit into a
2$\times $2 matrix which we denote by $ \Sigma _ {\rm R } $. Then the
rate can be written as
\begin{align} 
   \label{rateWeyl} 
   \frac{\text{d}\Gamma}{\text{d}^3 k} =
   \frac{2}{(2\pi)^3 k^0} f _ { \rm F } (k^0) \operatorname{Tr} \left[
     \overline \sigma  \cdot k  \operatorname {Im}  \Sigma_{\rm R, ret} (k) \right]
   ,
\end{align}
where $  \overline \sigma ^ 0 = \mathbbm{1} $,  $  \overline { \vec  \sigma  }
= - \vec \sigma  $. 

\section{Production  by collinear emission}      
\label{sc:LPM}

\subsection{Production processes}
\label{sc:processes}

Without any SM interactions the imaginary part of the 
self-energy is obtained by cutting  the   diagram
in Fig.~\ref{fg:2point}, with the lines in the loop being tree-level propagators. 
The corresponding production mechanism is  the 
inverse decay of the massive Majorana neutrino, $ \varphi  \ell \to
N $.

The typical momentum of
a particle in the hot plasma is $O(T)$.
We refer to these momenta as 'hard'. In the high-temperature
regime, when $ M _ N \lsim  g T $, one cannot neglect 
the modification of the dispersion relation of 
SM particles, which is caused by the interactions 
with other particles in the hot plasma.
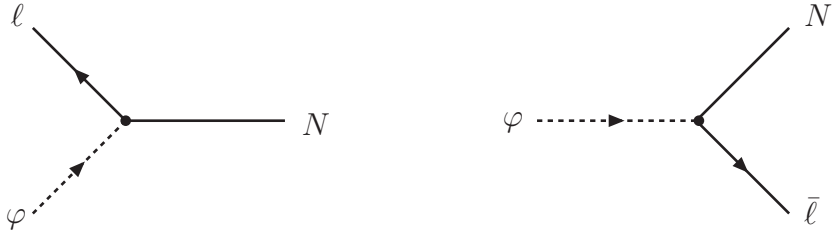
\begin{figure}[!t] 
\centering
  \begin{picture}(0,100) (200,-2)
    \SetWidth{1.0}

  \DashArrowLine(250,50)(310,50){2}
  \Line(310,50)(345,85)
  \ArrowLine(310,50)(345,15)
  \Vertex(311,50){2}
  \Text(352,86)[lb]{\normalsize{$N$}}
  \Text(352,11)[lb]{\normalsize{$\bar \ell$}}
  \Text(238,46)[lb]{\normalsize{$\varphi$}}
  \Line(95,50)(155,50)
  \DashArrowLine(60,15)(95,50){2}
  \ArrowLine(95,50)(60,85)
  \Vertex(95,50){2}
  \Text(162,45)[lb]{\normalsize{$N$}}
  \Text(50,9)[lb]{\normalsize{$\varphi$}}
  \Text(52,86)[lb]{\normalsize{$\ell$}}
\end{picture}
\caption{Decay and inverse decay processes that contribute to the production 
of Majorana neutrinos.} 
\label{fg:decay}
\end{figure}
The dispersion relation for hard particles 
can be written as $ p ^ 2 = m ^ 2 $, where $ m $ is the 
so-called asymptotic mass~\cite{Weldon_fermion}.~\footnote{It is
  oftentimes referred to as $ m_\infty  $.} 
It is given by the self-energy of a hard particle with
(nearly) light-like momentum, and the dominant contribution is due to hard loop momenta.
Its values for the Higgs and the lepton doublets
are given by \footnote{For scalars  the asymptotic  
   mass is the same as the thermal self-energy computed for vanishing
  momentum, which enters the finite temperature effective potential for the
  scalar field. It also 
  equals the frequency of scalar
  field oscillations with zero $ \vec p $. For fermions, however, 
the asymptotic mass is larger than the oscillation frequency for vanishing $ \vec p $
by a factor $ \sqrt{ 2 } $ \cite{Weldon_fermion}.}
\begin{align}  \label{asymptotic_masses}
   m_\varphi ^2 &=  \frac{1}{16} \left (  3g ^2 + g' {}^2 + 4 h_t ^2 +
     8 \lambda   \right ) T ^ 2          \, ,       
   \nonumber \\
   \quad m_\ell ^2 &=  \frac{1}{16} \left (  3g ^2  + g' {} ^2 \right ) T ^ 2
   .
\end{align} 
Note that the gauge field contributions to the asymptotic masses for 
Higgs and leptons
are equal (cf.\ Refs.~\cite{smilga}). However, the Higgs also
receives important 
contributions from the Yukawa interaction with the top quark and from the
Higgs self-interaction, so that $ m _ \varphi  > m _ \ell $. 
All other contributions can be neglected due to the
smallness of the corresponding coupling constants. 
Furthermore, the thermal mass of the Majorana neutrinos can be neglected, 
so that $ M _ N $ has a temperature-independent value. 
Therefore, the
Higgs and the leptons become ``heavier'' than the Majorana neutrinos at high $
T $. 

When $ m _ \varphi  + m _ \ell > M _ N $ 
 the inverse decay is kinematically forbidden. At even higher
temperature, when  $ m _ \varphi  - m _ \ell > M _ N $,  
the phase space for 
the decay $ \varphi  \to \bar \ell N $ of the Higgs boson opens up 
(see Fig.~\ref{fg:decay}). 
The  decay due to thermal masses has been considered as 
a mechanism for Majorana neutrino production 
previously~\cite{Giudice,Kiessig}.~\footnote{In Refs.~\cite{Giudice,Kiessig} 
the oscillation frequency for vanishing $ \vec p $  was used instead of the 
asymptotic mass for fermions.}

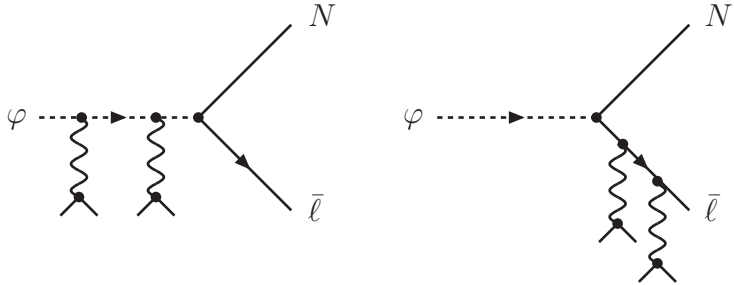
\begin{figure}[!t]
\centering
  \begin{picture}(-17,70)(205,-5)
    \SetWidth{1.0}

  \DashArrowLine(80,50)(140,50){2}
  \Line(140,50)(175,85)
  \ArrowLine(140,50)(175,15)
  \Photon(95,50)(95,20){3}{3}
  \Line(88,13)(95,20)
  \Line(102,13)(95,20)
  \Vertex(95,20){2}
  \Photon(124,50)(124,20){3}{3}
  \Line(117,13)(124,20)
  \Line(131,13)(124,20)
  \Vertex(124,20){2}
  \Vertex(96,50){2}
  \Vertex(124,50){2}
  \Vertex(140,50){2}
  \Text(182,85)[lb]{\normalsize{$N$}}
  \Text(182,11)[lb]{\normalsize{$\bar \ell$}}
  \Text(68,46)[lb]{\normalsize{$\varphi$}}
  \DashArrowLine(230,50)(290,50){2}
  \Line(290,50)(325,85)
  \ArrowLine(290,50)(325,15)
  \Photon(298,10)(298,40){3}{3}
  \Line(291,3)(298,10)
  \Line(305,3)(298,10)
  \Vertex(298,10){2}
  \Photon(313,-5)(313,25){3}{3}
  \Line(306,-12)(313,-5)
  \Line(320,-12)(313,-5)
  \Vertex(313,-5){2}
  \Vertex(300,40){2}
  \Vertex(313,26){2}
  \Vertex(290,50){2}
  \Text(332,85)[lb]{\normalsize{$N$}}
  \Text(332,11)[lb]{\normalsize{$\bar \ell$}}
  \Text(218,46)[lb]{\normalsize{$\varphi$}}
\end{picture}
\caption{Example for two processes whose interference needs to 
  be taken into account in a consistent leading order treatment.
}
\label{fg:interference}
\end{figure}
When $ M _ N \lesssim g T $, the characteristic feature of both processes
described above is that the momenta of the particles are nearly
\emph{collinear} because the relevant masses are small. The angles
between the particle momenta are of order $ g $.   We will show that 
multiple scattering mediated by soft gauge bosons contributes already at
leading order, similar to the thermal production of 
on-shell photons~\cite{AurencheLPM,arnoldPhoton}. The additional couplings
are compensated by nearly on-shell propagators. The lepton
and the Higgs boson undergo an arbitrary number of scatterings off
soft gauge bosons during the 'emission' of the Majorana neutrino which
still feels the presence of its 'source'. In order to take this
phenomenon into account, one has to resum an infinite set of diagrams,
like in~\cite{arnoldPhoton,BesakBodeker} for the production of
photons from a quark-gluon plasma. This will be dealt with
in the rest of Sec.~\ref{sc:LPM}.
\subsection{Momentum scales}
\label{sc:kinematics}

For the processes that we  consider the
4-momenta
of the emitting and of the emitted particles are
almost collinear and close to the light-cone. The 3-momenta
point in approximately the same direction, which we represent by the  3-vector $
\vec v $ with $ \vec v ^ 2 =1$. The components parallel to $ \vec v
$ are denoted by
\begin{align} 
    p _ \| \equiv \vec p \cdot  \vec v
    \label{pparallel} 
    .
\end{align} 
We further define the light-like vector  $ v \equiv 
( 1,  {\vec v }  ) $.

One has to account for three distinct momentum
scales: 
\begin{enumerate}
\item The emitting  particles 
and the emitted particle both have  $ p _ \| \sim T$, which is our hard scale.
\item All 3-momenta perpendicular to  $ \vec v
$ are soft, $ \vec p _\perp \sim g T $. Furthermore, all momentum components
 of the exchanged gauge bosons are
soft, 
$ q _ \mu  \sim g T $.
\item Finally, all 4-momenta $ k $ satisfy
$ v \cdot k = ( k _ 0 - k _ \| ) \sim g ^ 2 T $.  
\end{enumerate}

The hard loop momenta have $ k ^ 2 \sim g ^ 2 T^2 $. Therefore the
propagators are sensitive to the asymptotic mass $ m \sim g
T $.

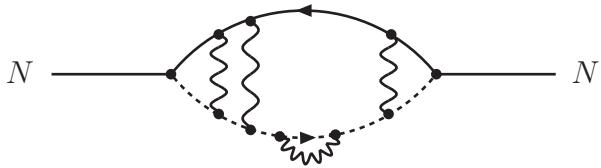
\begin{figure}[!t]
\centering
  \begin{picture}(-27,40)(205,-80)
    \SetWidth{1.0}
  \Line(105,-50)(150,-50)
  \Line(250,-50)(295,-50)
  \ArrowArc(200,-90)(64,38,142)
  \DashArrowArc(200,-10)(64,220,-37){2}
  \Photon(168,-35)(168,-64){3}{3}
  \Photon(180,-30)(180,-70){3}{3}
  \Photon(232,-35)(232,-65){3}{3}

  \PhotonArc(202,-72)(10,180,360){2}{7}
  \Vertex(150,-50){2}
  \Vertex(250,-50){2}
  \Vertex(168,-35){2}
  \Vertex(168,-65){2}
  \Vertex(180,-30){2}
  \Vertex(180,-71){2}
  \Vertex(233,-35){2}
  \Vertex(232,-65){2}
  \Vertex(191,-73.5){2}
  \Vertex(212,-72.8){2}
  \Text(88,-53.5)[lb]{\normalsize{$N$}}
  \Text(302,-53.5)[lb]{\normalsize{$N$}}
\end{picture}
\caption{Example for a self-energy diagram whose imaginary part contributes to
  the leading order production rate.
}
\label{fg:ladder}
\end{figure}
\subsection{Strategy of the calculation}
\label{sc:strategy}

We now turn to computing the production rate due the processes discussed in 
Sec.~\ref{sc:processes}.  
They involve interference terms with an
arbitrary number of soft  SU(2) and U(1)  gauge bosons 
and the corresponding self-energy needed for \eqref{rate} is given by
the sum of all ladder diagrams of the type shown in Fig.~\ref{fg:ladder}.
We use
the approach of Ref.~\cite{BesakBodeker} to obtain an integral equation
for the self-energy that we can subsequently solve numerically and get
the contribution to the production rate.
Before going into the details, let us recall the strategy of the 
calculation~\cite{BesakBodeker}:
\begin{enumerate}
\item Integrate out the hard field modes. This generates the
  asymptotic masses for hard particles near the light-cone, i.e. with
  $ p ^ 2 \sim g ^ 2 T ^ 2 $.  Furthermore, one obtains the Hard
  Thermal Loop (HTL) effective action for soft fields, in particular
  the gauge fields. At leading order no thermal width for the hard
  particles is generated in this step. 
  The width due to hard-hard interaction is of order $ g ^ 4 T $ and
  can be neglected. The leading order width is due to soft gauge
  interactions and enters only at a later stage.
\item Consider 1-loop diagrams with 2 external Majorana neutrinos and with
  an arbitrary number of external soft gauge bosons, in the
  limit that the loop momentum and the hard external momenta are nearly
  collinear. Expand the propagators and vertices in the coupling and
  keep only the leading order terms. First compute the 2-point function
  without any gauge boson explicitly. Then set up a recursion relation
  that relates a given $n$-point function to $(n - 1)$-point functions
  where one of the gauge bosons has been removed.  This recursion
  relation is most easi\-ly formulated in terms of a `current' which is
  induced by background Majorana neutrino and gauge fields.
\item Finally, integrate out the soft gauge boson background. The
  gauge bosons then only appear in self-energy insertions (thereby
  generating the thermal width for the lepton and the Higgs boson) and
  as rungs in the ladder diagrams (see
  Fig.~\ref{fg:ladder}).~\footnote{At leading order, only diagrams of
    the form shown in Fig.~\ref{fg:ladder} are generated. No diagrams
    with crossed ladder rungs or nested loops
    occur~\cite{arnoldPhoton,BesakBodeker}.}  The new current
  satisfies an integral equation that is straightforwardly obtained
  from the one found in step 2. By stripping off the external field we
  obtain an integral equation for the desired self-energy.
\end{enumerate}
In the remainder of Sec.~\ref{sc:LPM}, we will provide all the relevant
details to arrive at the final result \eqref{ratepsif} for
the production rate of Majorana neutrinos, following the strategy 
outlined above. 

\subsection{2-point function}\label{sc:2point}

In this section we explicitly compute the 2-point function of the Majorana 
neutrinos shown in Fig.~\ref{fg:2point},  ignoring  any interactions,
except that we use the asymptotic 
masses. 
It turns out that 
all $n$-point functions  with additional  $(n - 2)$ soft gauge bosons 
can be obtained from $ (n - 1) $-point
functions through a simple recursion relation.
All simplifications of propagators and vertices which are needed for
the general case already appear in the calculation of  the 2-point function. 
  
Let us first have a look at the propagators. Keeping only the
leading order $ g ^ 2 T ^ 2 $  terms in the denominator,
the scalar propagator with mass $ m _ \varphi   $ can be written as~\cite{BesakBodeker}
\begin{align} 
  \Delta  ( p ) \equiv \frac{ -1}{ p ^ 2 - m  _ \varphi   ^ 2 } 
  \simeq     \frac{ D _ \varphi  (p) }{ 2p _ \| } 
   \label{propagator}
\end{align} 
with 
\begin{align} 
   D _ a ( p ) 
   \equiv 
   \frac{ -1}{ v \cdot p  
               - (   \vec p _\perp ^ 2 + m  _ a ^ 2 ) / ( 2 p _ \| ) 
              } 
    \label{D}
    .
\end{align} 
Now consider a spin 1/2 propagator. 
We consider only left-handed fermions propagating in the loop. Thus we can work with
2-component Weyl spinors. One can approximate
\begin{eqnarray} 
  S ( p )    \simeq     \frac{ D _ \ell ( p ) }{ 2 p _ \|} 
  \sigma  \cdot \widetilde p 
\end{eqnarray} 
with
\begin{align} 
  \widetilde {p } \equiv   p  
  - \frac{m_{\ell } ^2}{2p_\parallel} u,
\end{align} 
where $ u \equiv  (1, \vec 0) $ is the 4-velocity of the plasma.
The matrix $ \sigma  \cdot \widetilde p $ contains terms of order $ T $, $ g T
$ and $ g ^ 2 T $; higher order terms have been neglected. It is necessary
to keep the $ g ^ 2 T $ terms because they give a leading order contribution
to the rate in Eq.~(\ref{rate}). 

If the fermion momentum is on-shell, $ p ^ 2 = m _ \ell ^ 2 $, then
$ \widetilde { p} $ is light-like  
up to higher order terms which we have
neglected. Therefore, if we evaluate the loop integral 
in the imaginary-time formalism and take the imaginary part, the fermion
propagator is on-shell, and we can treat  $ \widetilde { p} $ as a light-like
4-vector. This allows us to write
\begin{eqnarray} 
   \sigma  \cdot \widetilde p  
   = 2 p _ \| \eta  \left ( \widetilde p \right ) \eta ^\dagger  ( \widetilde
   p ) 
   , 
\end{eqnarray}
with a 2-component Weyl-spinor $ \eta  ( \widetilde p ) $. Then the
fermion propagator becomes 
\begin{eqnarray} 
  S ( p ) = \eta  \left ( \widetilde p \right ) \eta ^\dagger  ( \widetilde p
  ) 
   D _ \ell ( p ).
   \label{S}
\end{eqnarray} 
We choose the $ 3 $-axis in the direction of $ \vec v $ and expand
 $ \eta  ( \widetilde p  ) $ in powers of $ g $, 
\begin{align}
  \eta  
  = \eta  _ 0 
  + \eta  _ 1 
  + O (g ^2) 
  \label{eta}
\end{align} 
with
\begin{align}
   \eta _ 0
   = \left ( 
   \begin{array}{c}
     0\\1
   \end{array}
 \right ) 
 , \qquad 
 \eta _ 1
 = - \frac{ \vec \sigma  \cdot \vec p _\perp } { 2 p _ \| } 
 \eta  _ 0 
 .
 \label{eta0}
\end{align}
The lower component is $ O ( 1 ) $ and the upper component is $ O ( g )
$. Since we only need the leading order for each component of $ \eta  $,
which is at most of order $ g $,  
we can write $ \eta  ( p ) $
instead of $ \eta  ( \widetilde p ) $ because the difference of $ p $ and $
\widetilde p $ is of order $ g ^ 2 T $.

We use the partial fractioning 
\begin{align} 
  D _ \ell ( p ) D _ \varphi  ( p - k ) 
  =  \frac{ 1 } { \epsilon  
    ( k ,\vec p ) } 
  \left [ D _ \ell ( p ) - D _ \varphi  ( p - k ) \right ]
  \label{fractioning}
  ,
\end{align} 
where
\begin{align}\label{epsilon}
  \epsilon  (k,\vec p) 
  \equiv v \cdot k + \frac{(\vec p_\perp - \vec k_\perp)^2 + m_\varphi  
  ^2}{2(p_\parallel - k_\parallel)} - \frac{\vec p_\perp ^2 + m
  _ \ell 
  ^2}{2p_\parallel}
\end{align}
is the difference of the energy poles of the two propagators. Then it is
straightforward to compute the self-energy.
It will turn out to be convenient to 
write it in the form 
\begin{align} 
  \label{reduced}
   \Sigma  _ { \rm R } (k) = |h|^2 \int \frac{\text{d}^3 p}{(2\pi)^3} 
  \eta  ( 
      p ) 
   \widehat \Sigma (k,\vec p)
   .
\end{align} 
It is proportional to $|h|^2 \equiv \sum_j
|h _{1j}|^2$   where the sum goes over the families of leptons.
The gauge
interactions which will be included are the same for each lepton family, so
that this factor can always be extracted. 
Our result for the  'reduced self-energy' $\widehat
\Sigma  (k,\vec p)$ is
\begin{align}
  \widehat \Sigma (k,\vec p) = - 
  \frac{ d ( r )    \mathcal{F}(k_\parallel,p_\parallel) } 
  {2  \epsilon (k,\vec p)  } 
  \frac{ \eta  ^\dagger  ( p ) } { p _ \| - k _  \| } 
  \label{2point} 
  ,
\end{align}
where $ d ( r ) = 2$ denotes the dimension of the gauge group representation for 
lepton and Higgs, and 
\begin{align}
  \mathcal{F}(k_\parallel,p_\parallel)
  \equiv  f _ { \rm F }  (p_\parallel) +  f _ { \rm B } 
  (p_\parallel - k_\parallel)
  .
\end{align}
From Eq.\ (\ref{2point})  one then obtains the rate due to the decay of the
Higgs 
or from the inverse decay of the Majorana neutrino, with the diagrams shown in
Fig.~\ref{fg:decay}. Explicit results can be found in Sec.~\ref{sc:treelevel}.

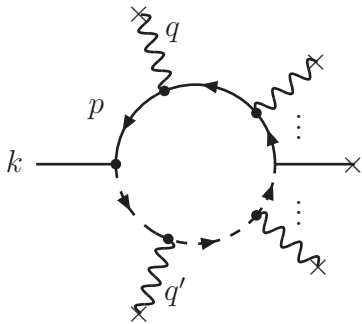
\begin{figure}[t] 
  \centering

\begin{picture}(50,90)(130,-0)

\SetWidth{1}

\Text(80,61.5)[]{ $k$}

\Line(120. , 60.)( 90., 60.)
\Vertex(120. ,60. ){2}
\ArrowArc(150,60)(30,120,180)
\Vertex(138.343 ,87.6427 ){2}
\Text(140,110 )[]{ $ q $}
\Photon(129.479,116.382)(139.739,88.1908){3}{4}
\Text(131,117)[]{$ \times $ }
\ArrowArc(150,60)(30,40,120)
\Vertex(172.98 , 79.28 ){2}

\Photon(172.98 , 79.28 )(195.96 , 98.57 ){3}{3.5}
\Text(198 , 99 )[]{$ \times $ } 
\ArrowArc(150,60)(30,0,40)
\DashArrowArc(150,60)(30,320,0){5}
\Vertex( 172.98, 40.7){2}

\Photon( 173, 40.7)(195.96 , 21.4 ){3}{3.8}
\Text(199 , 21.4 )[]{$ \times $ } 
\DashArrowArc(150,60)(30,240,320){5}
\Vertex( 139.739, 31.8092){2}
\Text(141 , 12)[]{ $ q'  $}
\Photon(139.739 ,31.8092 )(129.479 , 3.61844){3}{4}
\Text(131 , 3.61844)[]{$ \times $ } 
\DashArrowArc(150,60)(30,180,240){5}

\Text(191.575 , 77.2208)[]{$ \vdots $ }

\Text(191.575 , 44.7792)[]{$ \vdots $ }

\Line(180,60)(210,60)
\Text(212,60)[]{$ \times $ }

\Text(111,81)[]{ $p$}

\end{picture}

\caption{1-loop diagrams with soft external gauge field lines. 
  Only the 2-point function Fig.~\ref{fg:2point}  
  needs to be calculated explicitly.
  The $ n $-point functions with $n > 2$ are related to the $ ( n -1 )  $-point functions 
  by a recursion relation. }
\label{fg:npoint}
\end{figure}
\subsection{Self-energy in soft gauge field background}
\label{sc:background}

Now we consider diagrams with 2 external $ N $'s and, in addition, 
$ (n - 2) $ external gauge field lines. These diagrams can  be recursively 
related to 
diagrams with $ (n - 3) $ gauge fields. To obtain
this relation we pick out one vertex, the leftmost one in
Fig.~\ref{fg:npoint}, with incoming momentum $ k $. 
As for the 2-point function we can use the approximations 
\eqref{D} and \eqref{S} for the propagators. 
After the partial fractioning (\ref{fractioning}) 
this diagram is proportional to the difference of the diagrams in which either
the external line with  momentum $ q $ or the one  
with momentum $ q ' $ is omitted. 

It is convenient to work with 
the generating functional of the diagrams rather than with the diagrams 
themselves. We attach background fields to the external lines, except to 
the one on the left. That means that we consider the first derivative of 
the generating functional. It can be interpreted as a 'current' which 
is generated by the presence of the background fields. 

Here we simplify the treatment of Ref.~\cite{BesakBodeker}, where
first diagrams were considered and then the generating functional was
built from them.  Instead we start directly with the current.  We
define $ \widehat J ( k , \vec p ) $ as the sum of all diagrams in the
background of external Majorana neutrino $ N $ and of gauge fields $ A
_ \mu ^ A $, but without the
integral over the 3-momentum $ \vec p $, without the trace over the SU(2)
indices, and without  the vertex factor $  \eta  ( p ) $ 
(cf.\  Eq.~(\ref{reduced})).   
Each lepton family gives the same contribution.
Therefore we can also pull out the factor $ | h | ^ 2 $ as in
Eq.~(\ref{reduced}). The current without gauge fields, which
corresponds to our result from Sec.~\ref{sc:2point}, is
\begin{align}
   {\rm tr} \, \widehat  J ( k, \vec p ) =  \widehat  \Sigma ( k , \vec p ) 
    N ( k ) 
    \label{current}
    .
\end{align} 
Here `tr' refers to the trace over SU(2) indices. 

For computing higher $ n $-point functions it is furthermore convenient to 
associate the
factor $ ( 2 p _ \| ) ^{ -1 } $  in the scalar propagator $ \Delta  ( p ) $ 
(see  Eq.~(\ref{propagator})) with the vertex in
the direction of $ p $, 
rather than with the propagator itself. Then the vertex of two Higgs lines
with one gauge field with soft momentum $ q $ is 
\begin{align} 
  \frac{ 1 } { 2 p _ \| } ( 2 p - q ) ^  \mu 
  = v ^ \mu  + O ( g ) .
  \label{scalarvertex}
\end{align} 
On the fermion lines 
we associate the spinors $ \eta  ( p ) $ and $
\eta  ^\dagger ( p ) $ not with the propagator, but with the vertices to which
the propagator is attached.  For a fermion-gauge field vertex one
therefore obtains a factor
\begin{align} 
   \eta  ^\dagger ( p - q ) \bar \sigma  ^ \mu  \eta
   ( p ) 
   = v ^ \mu  + O ( g )
   \label{fermionvertex}
.
\end{align} 
From Eqs.~(\ref{scalarvertex}) and (\ref{fermionvertex}) we only need the
leading order contributions, which are equal. 

For the two propagators 
which
are connected to the left vertex in Fig.~\ref{fg:npoint}
 we use the same  approximations \eqref{D} and \eqref{S}
as for the computation of the 2-point function, and we use the
partial fractioning~(\ref{fractioning}). The resulting two terms 
are then proportional  to diagrams in which either the pro\-pa\-gator with
momentum $ p $ or the one with momentum $ p - k $ has been omitted from
Fig.~\ref{fg:npoint}. Thus, if one also leaves out the vertex factors and
the background gauge fields 
connected to these propagators, each of the two terms gives a
 ($ n -1 $)-point function.
For the second term in Eq.~(\ref{fractioning}) we obtain a contribution in 
which the propagator with momentum $ p $ has been
omitted. 
We perform a shift in the summation variable, $ p ^ 0 \to p ^ 0 
+ q ^ 0 $. Then the remaining propagators are the same which appear in the 
$ ( n - 1 ) $-point function, but with the loop 3-momentum $ \vec p $ replaced by 
$ \vec p - \vec q $. 
Therefore  we can write $ \widehat J  
 ( k, \vec p ) $ 
in terms of the difference of 
$ \widehat J $ times a gauge field and of a gauge field times 
$ \widehat  J $, 
\begin{align}
  \epsilon    ( k, \vec p ) \widehat J ( k, \vec p ) 
  =   & \, 
  - \frac12 \mathcal{F}(k_\parallel,p_\parallel) 
  \frac{ \eta  ^\dagger  (p ) } { p _ \| - k _ \| } 
    N  ( k )  \nonumber \\ &
    {} + \int _ q    
    \left [  \widehat J ( k - q , \vec p )   v \cdot A( q ) 
    - v \cdot A ( q )  
     \widehat J ( k - q , \vec p - \vec q )     \right ]  
         \label{72}
         .
\end{align} 
In the imaginary-time formalism
\begin{align} 
  \int_{q} \equiv T\sum_{q^0 = i \omega  }\int \frac{\text{d}^3 q}{(2\pi)^3} 
  ,
\end{align}
where $  \omega  $ is a bosonic Matsubara frequency.
Furthermore, $ A _ \mu  \equiv A _ \mu  ^ A T ^ A $ contains both SU(2) and U(1) gauge fields.
Note that we have included the gauge couplings in $ A _ \mu  ^ A $. They will appear in 
the gauge field propagators.   
Eq.~(\ref{72}) is of the same form as Eq.~(21) in~\cite{BesakBodeker}, 
only the inhomogeneous term on the RHS is different. 

\subsection{Integrating out the soft gauge bosons}
\label{sc:integrating}

Now we use Eq.~\eqref{72} to integrate out the soft gauge fields. This
will give us the $ N $ self-energy including soft gauge interactions.  
To see how it works, we write Eq.~(\ref{72}) schematically as 
$ \widehat J = N + A \widehat J $, leaving out all terms and all factors which
are not essential for this purpose.  We iterate this equation once, which
gives $  \widehat J = N + A ( N + A \widehat J )$. Now
we integrate this expression over the gauge fields. The term linear in $ A $
drops out and we obtain $  \left \langle \widehat J \right  \rangle = N
+  \left \langle A A \widehat J \right \rangle $, where $
\langle \cdots \rangle $ denotes the path integral over the soft $ A $-field.  In
\cite{BesakBodeker} we have shown that at leading order $ \left \langle  A A
  \widehat J \right \rangle = \left \langle \vphantom{\widehat J} A A \right \rangle \left \langle
  \widehat J \right \rangle $. In this way we obtain a closed equation for $
\left \langle \widehat J \right \rangle $.

Now we can become more explicit. We iterate Eq.~(\ref{72}) once and we 
take the trace over the SU(2) indices. This allows us to bring all terms in
the same order,
\begin{align} 
\lefteqn{
  \epsilon     ( k, \vec p ) \,\, {\rm tr} \left\langle  \widehat J (k,\vec p)
\right\rangle 
   = -
   \frac{d(r)}{2}    \mathcal{F}(k_\parallel,p_\parallel) 
  \frac{ \eta  ^\dagger ( p ) } { p _ \| - k _ \| } N ( k )  } \nonumber \\
  &&  {} +  2 
   \int _ q \int _ {q'}
   \frac{1}{v\cdot(k - q)} {\rm tr} 
   \left[ 
     \left \langle v \cdot A ( q ) v \cdot A ( q ' ) \right \rangle 
       \left\langle 
         \widehat J (k,\vec p)  -  \widehat J (k,p_\parallel,\vec
       p_\perp - \vec q _\perp ) \right\rangle \right]  \hfill 
   \label{74.1}
   .
\end{align} 
Inside the integral  we have approximated  $ \epsilon(k - q,
\vec p ) \simeq  \epsilon(k - q,
\vec p  - \vec q ) 
\simeq v  \cdot (  k - q  ) $, i.e. we have neglected the terms containing transverse
momenta  and thermal masses, even though they are of the same order as the
term we kept. This is possible because the gauge field 
propagator separately depends on $ q _ 0 \sim gT $ and $ q _ \| \sim gT$, and not on their difference
$ v \cdot q \sim g^2 T $. Therefore the terms we have omitted only contribute to a higher order
shift of the integration variable $ q _ \| $.

For the SU(2)$ \times $U(1) gauge fields one has at leading order
\begin{align} 
   \left\langle A_\mu (q) A_\nu (q') \right\rangle = 
   \tilde \delta  ( q + q' ) 
   \left [ 
     C _ 2 ( r ) g ^ 2 \Delta_{\mu \nu}(q) 
     + y _ \ell ^ 2 g'  {}^ 2  \Delta  '      _{\mu \nu}(q)
   \right ] 
\label{75}
\end{align} 
where in the imaginary-time formalism 
\begin{align} 
   \tilde \delta  ( q + q' ) \equiv 
   T ^ {-1} \delta  _ { q _  0+ q ' _ 0 , 0} ( 2 \pi  ) ^ 3 \delta  ( \vec q +
   \vec q ' ) 
   .
\end{align} 
Furthermore,  $C_2(r)$ is
the quadratic Casimir operator which for the fundamental re\-pre\-sentation of
SU(2) equals 3/4, and $ y _ \ell =
-1/2 $ is the lepton hypercharge. Finally, $  \Delta_{\mu \nu} $ and $
\Delta_{\mu \nu} ' $ are the HTL resummed propagators of the
SU(2) and U(1) gauge fields. 

On the RHS of  Eq.~(\ref{74.1}) the unknown function $ \widehat J $ does not
depend on $ q _ 0 $ or $ q _ \| $. Therefore these integrals can be performed.
One encounters the expression
\begin{eqnarray} 
  I ( k , \vec q _\perp ) 
  \equiv  T\sum _ { q ^  0 = i \omega  } \int \frac{  \text{d} q _ \| } {  2 \pi   } 
    \,\,
  \frac{  v^\mu
    v^\nu \Delta_{\mu \nu}(q)  }{v\cdot(k - q)  }
   \label{beforesumrule}
   , 
\end{eqnarray}
and similarly with $\Delta'_{\mu \nu}$.
Here $k^0$ is purely imaginary, $ k ^ 0 = i\tilde \omega  $, where $\tilde
\omega   $  is a
fermionic Matsubara frequency. After performing the sum over
the bosonic Matsubara frequencies $ \omega  $,
one has to analytically continue to
$ k ^ 0 + i 0^ +   $, where  $ k ^ 0 $ is now real. 
Using  standard results for the HTL resummed propagators~\cite{htl}
and the sum rule of Ref.~\cite{Aurenche1} one obtains
\begin{align} 
  I ( k ^ 0 \pm i 0 ^ +  , \vec k , \vec q _\perp ) 
   \simeq 
   \mp \frac{ i } { 2 } T 
   \left ( 
     \frac{ 1 } { \vec q _\perp ^ 2 }   - \frac{ 1 } { \vec q _\perp ^ 2 +
       m^2_{\rm D} }
   \right ) 
   \label{sumrule}
\end{align} 
where $m_{\rm D}$ is a Debye mass.  In the SM the Debye
masses of the SU(2) and U(1) gauge bosons $ m_{\rm D} $ and $ m_{\rm
  D} ' $ are given by~\cite{carrington}
\begin{align} 
  m_{\rm D} ^ 2= \frac{11}{6} g^2 T^2, \quad m_{\rm D} ' {} ^ 2
  = \frac{11}{6} g'^2 T^2 
  .
\end{align} 
The RHS of Eq.~(\ref{sumrule}) no longer depends on $ k $. The only $ k $-dependence of $ I
$ is through the sign of the imaginary part of the frequency.

Now we strip off the background $ N $-field from $  \widehat J $
which gives $ \widehat \Sigma
 $ (see Eq.~(\ref{current})). The equation for 
$ \widehat \Sigma  (k,\vec p)    $ then reads
\begin{align} 
  i \epsilon   (k,\vec p)  \widehat \Sigma  (k,\vec p)   
  =& \, - \frac{ i }{2} d ( r )\mathcal{F}(k_\parallel,p_\parallel) 
  \frac{ \eta  ^\dagger ( p ) } { p _ \| - k _ \| } 
  \hfill 
  \nonumber  \\
  & +     \int \frac{\text{d}^2 q _\perp }{(2\pi)^2} \mathcal{C}(\vec
  q_\perp) 
  \left[ \widehat \Sigma (k,\vec p) 
    - \widehat  \Sigma  (k,p_\parallel,\vec p_\perp - \vec q _\perp ) \right ]
  \vphantom{\int} \hfill  
  \label{inteq}
\end{align}
with the kernel
\begin{align} 
  {\cal C }  ( \vec q _\perp ) \equiv 
  T \left [ C _ 2 ( r ) g ^ 2 
    \left ( 
      \frac{ 1 } { \vec q _\perp ^ 2 }   
      - \frac{ 1 } { \vec q _\perp ^ 2 + m^2_{\rm D} }
    \right ) 
    + y _ \ell ^ 2 g ' {} ^ 2 
    \left ( 
      \frac{ 1 } { \vec q _\perp ^ 2 }   
      - \frac{ 1 } { \vec q _\perp ^ 2 + m_{\rm D}' {}^2 }
    \right ) 
  \right ]    . 
\end{align}
The integral equation \eqref{inteq} is of the same form as the one for the 
production of photons~\cite{BesakBodeker}, only the inhomogeneous term is different. Both 
$\epsilon(k,\vec p) \sim g^2 T$ and $\text{d}^2 q_\perp 
\mathcal{C}(\vec q_\perp) \sim g^2 T$. Therefore the integral term 
which resums the multiple interactions with an arbitrary number of 
soft gauge bosons is of the same order as the term which corresponds 
to the tree-level (inverse) decay.

\subsection{Production rate of Majorana neutrinos}
\label{sc:application}

To compute the trace over spinor indices in Eq.~(\ref{rateWeyl}) we expand
\begin{align} 
  \overline \sigma  \cdot k = 
  2 k _ \| \chi  _ 0 \chi  _ 0 ^\dagger 
  + \frac{ M _ N ^ 2 } { 2 k _ \| } \eta  _ 0  \eta  _ 0 ^\dagger 
  + O ( g ^ 3 T)  
  \label{sigmak}
\end{align} 
where $ \eta  _ 0 $ is given by Eq.~(\ref{eta0}), and 
\begin{eqnarray} 
  \label{chi0}
     \chi _ 0
   = \left ( 
   \begin{array}{c}
     1\\0
   \end{array}
 \right )
 . 
\end{eqnarray} 
Since the first term in Eq.~(\ref{sigmak}) is $ O ( T ) $ and the second is
$ O ( g ^ 2 T ) $, 
both terms in Eq.~(\ref{eta}),  and consequently both components of $ \widehat
\Sigma  (k,\vec p)   $, contribute to the leading order production rate.

Now, if one has the solutions to the equations
\begin{align}  
  \label{inteq_forf} 
  i\epsilon(k, \vec p) \vec f ( \vec p _\perp ) 
  - 
  \int \frac{\text{d}^2 q_\perp}{(2\pi)^2} 
  \mathcal{C} (\vec q_\perp)  
  \left[ \vec f (\vec p_\perp) - \vec f (\vec p_\perp - \vec q_\perp) \right]
  & = 2 \vec p _\perp 
  \hfill 
  ,
  \\
  \label{inteq_forpsi} i\epsilon(k , \vec p)\psi(\vec p_\perp) 
  - 
  \int \frac{\text{d}^2 q_\perp}{(2\pi)^2} 
  \mathcal{C} (\vec q_\perp)  
  \left[ \psi(\vec p_\perp) - \psi(\vec p_\perp - \vec q_\perp) \right] 
   & = 1
   ,
\end{align}
which were obtained in~\cite{arnoldPhoton,AurencheDilepton} for transverse and
(virtual) longitudinal 
photon production, 
 then the
solution to Eq.~(\ref{inteq})  is given by 
\begin{align}\label{Identification}
   \widehat {  \Sigma}  (k,\vec p) 
   = - \frac{ i }{2} \, \frac{
     d(r)\mathcal{F}(k_\parallel,p_\parallel)}{p _ \| - k _ \|  }  
   \left (
     \begin{array}{c}
       - (  f ^ 1 + i f ^ 2 ) / p _ \|   \\
       \psi  
     \end{array}
   \right ) 
   .
\end{align}
Unless one ignores the interactions with soft gauge fields by  putting $ \cal
C $ equal to zero, 
 the imaginary part of $ k ^ 0 $ is no longer relevant in
 Eqs.~(\ref{inteq_forf}) and (\ref{inteq_forpsi}). Only its sign has entered by determining  the
 sign of $ \cal C $ through Eq.~(\ref{sumrule}). 

We may now choose $ \vec v $ in the direction of $ \vec k $, so that
$ \vec k _\perp = \vec 0 $, and $ k _ \| = | \vec k | \equiv k $.
Combining Eqs.\ \eqref{rateWeyl}, \eqref{eta0}, \eqref{reduced}, 
\eqref{sigmak}, \eqref{chi0},
\eqref{Identification}  and using
\begin{align} 
  f _ { \rm F } ( k ) [  f _ { \rm F } ( p ) + f _ { \rm B } ( p - k ) ] 
  = - f _ { \rm F } ( p )  f _ { \rm B } ( k - p ) 
\end{align} 
we obtain for the production rate 
\begin{align}
  \label{ratepsif}
      \frac{\text{d}\Gamma}{\text{d}^3 k} 
      = -\frac{d ( r ) |h|^2}{(2\pi)^3 2 k} 
       & 
      \int \!\! \frac{\text{d}^3 p}{(2\pi)^3} \frac{1}{k - p_\parallel} \hfill
      f_{\rm F} (p_\parallel) 
         f_{\rm B} (k - p _ \parallel )
        \operatorname{Re} 
      \left[ 
        \frac{ k } { 2 p _ \| ^ 2 } \vec p _\perp \cdot \vec f
        + \frac{M _ N ^ 2 }{k} \psi   
      \right] \hfill 
      . 
\end{align}
The integral equations (\ref{inteq_forf}) and
(\ref{inteq_forpsi}) for  $\vec f$ and $ \psi  $ can only be solved numerically
which is greatly simplified by transforming them via a
Fourier transformation into a boundary-value problem for a differential
equation as shown in appendix~\ref{app:Diffeq_LPM}.  

The Yukawa interaction couples Majorana neutrinos to SM leptons with
opposite chirality. In the massless limit where helicity equals
chirality the collinear emission is forbidden by angular momentum
conservation. This argument is not affected by the thermal masses for
fermions because they do not violate chiral symmetry.  The
contribution containing $ \vec f $ corresponds to helicity changing
processes, and therefore it vanishes when $ \vec p _\perp $ is zero.
$ \psi $ does not vanish in the collinear limit and corresponds to
helicity conserving processes. Therefore $ \psi $ vanishes in the
limit $ M _ N \to 0 $.

\subsection{Tree-level (inverse) decay}
\label{sc:treelevel}

In addition to solving the full integral equations \eqref{inteq_forf}
and \eqref{inteq_forpsi}, we want to explicitly calculate the
contribution from the tree-level processes shown in Fig.~\ref{fg:decay}
in order to study how strongly the multiple soft
scattering affects the production rate. 
This corresponds to neglecting the integral term in
\eqref{inteq} completely, and the equation becomes a purely algebraic
one. In this case one has to keep a small imaginary part of $ k ^ 0 $
in $ \epsilon ( k ,\vec p ) $. Taking the real part of the trace in
\eqref{ratepsif} and performing the integration over $\vec p_\perp$
yields
\begin{align}
  \label{decay_treelevel}
  \frac{\text{d}\Gamma ^{\rm  tree}}{\text{d}^3 k} =
  \frac{d ( r ) |h|^2}{ 4 (2\pi)^3 k ^2}  \int\limits _ { p _ - } ^ {p _ +}
  \frac{\text{d}p_\parallel}{2\pi} \frac{1}{|p_\parallel|} 
   f_{\rm F} (p_\parallel)  f_{\rm B}    (k - p_\parallel ) 
    \left[
    p_\parallel( M_N ^2 - m_\varphi ^2 ) - (k - p_\parallel )
    m_\ell ^2 \right].  
\end{align} 

The integration limits are determined by the
conditions that the energy of the lepton be positive and that the
particles be on-shell, which means that $\epsilon(k,\vec p) = 0$ has a
real solution for $| \vec p _\perp | $. For the two processes shown in Fig.\
\ref{fg:decay}, this gives the integration limits
\begin{align} 
  \label{integration_range_treelevel} 
  p _ { \pm } = \frac{X \pm \sqrt{Y}}{2M_N ^2} 
 \end{align}
where 
\begin{align}
  X \equiv & \, M_N ^2 + m_\ell ^2 - m_\varphi ^2  ,
  \nonumber \\
 Y \equiv & \, (m_\varphi + m_\ell + M_N)(m_\varphi - m_\ell +
M_N)(m_\varphi + m_\ell - M_N)(m_\varphi - m_\ell - M_N). 
\end{align}
It is easy to see that $Y \geq 0$ only if either $m_\varphi \geq M_N +
m_\ell$ (which corresponds to the decay of the Higgs boson) or $M_N
\geq m_\varphi + m_\ell$ (which corresponds to the inverse decay of
the Majorana neutrino). For $m_\varphi - m_\ell < M_N < m_\varphi +
m_\ell$ we therefore obtain $\text{d}\Gamma ^{\rm tree}/\text{d}^3 k
= 0$.

\section{Numerical results}
\label{sc:Numerical}

To obtain numerical results we have to specify several parameters. 
The mass $ M _ N $ of the Majorana neutrino and
its Yukawa couplings are unconstrained by
low-energy neutrino physics. We have chosen the exemplary value $M_N = 10^7$
GeV and we always plot our rates divided by $|h|^2 = \sum_j |h_{1j}|^2$.
The SM couplings are 
evaluated at the scale $\mu = 2\pi T$ using the 1-loop renormalization group 
equations~\cite{RG}. To determine the  Higgs self-coupling $\lambda$ 
we have assumed  $m_H = 150$~GeV for the  zero temperature Higgs mass. 
\begin{figure}[!h]
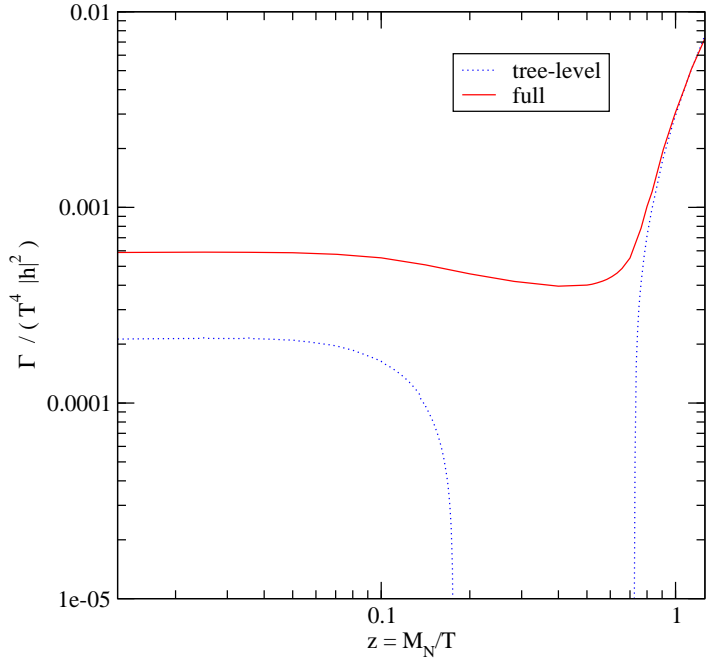

\centerbox{0.6}{collinearSum.eps}\vspace{0.5cm} 
\caption
{Number of produced Majorana neutrinos per unit time and unit volume 
 as a function of $ z \equiv  M _ N /T $. 
The dotted curve is the result without any soft gauge interactions. The
full line includes an arbitrary number of soft gauge interactions. 
\label{fg:rate}}
\end{figure}

First consider the integrated production rate $ \Gamma  =
\int \text{d} ^ 3 k ( \text{d}  \Gamma  /\text{d}^ 3 k  ) $ 
as a function of $ z = M _ N /T $. 
Our results are shown in Fig.~\ref{fg:rate}. 
At small $ z $ the 'tree-level rate' \eqref{decay_treelevel}
is due to the decay of Higgs bosons, while at large $ z $ it results from 
the inverse decay of heavy Majorana neutrinos.
At intermediate $ z $ none of these processes is kinematically allowed, and the rate
vanishes. It is remarkable that the full line which includes soft
gauge interactions is very smooth in the regions where the decay of the Higgs boson
becomes kinematically forbidden.

One can see that the full rate is larger than the tree-level rate by about a
factor 3 at small $ z $. It decreases only mildly when the tree-level processes
are forbidden. When the inverse decay process sets in, the difference
between tree-level rate and the complete rate goes to zero. This is expected, since
the collinear enhancement is a relativistic effect and it disappears when the
Majorana neutrinos become non-relativistic. 
One should emphasize that the strong enhancement caused by the soft gauge 
interactions does not signal a breakdown of perturbation theory, because
all contributions discussed above are leading order.

\begin{figure}
\centerbox{0.6}{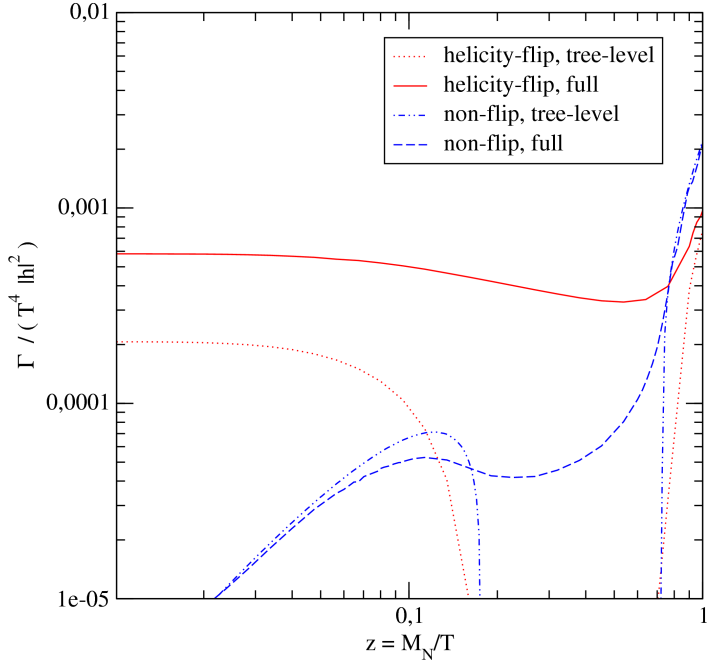} 
\caption
{Contributions to the production 
rate of Majorana neutrinos due to helicity changing 
and helicity conserving processes 
 as a function of $ z \equiv  M _ N /T $.
\label{fg:heli}}
\end{figure}

It is also interesting to consider the contribution due to helicity changing
and helicity conserving processes separately (cf.~the discussion at the end of
Sec.~\ref{sc:application}). The rate of helicity changing processes does not vanish in
the limit $ M _ N \to 0 $, and should  therefore  be dominant at small $ z $.   
The results are shown in Fig.~\ref{fg:heli}. We
clearly see that the helicity changing process dominates at high temperatures
and the helicity conserving process is relevant only at low temperatures, $T
\sim M_N
$, while it becomes negligible at $T \gg M_N $. This is true both for the
tree-level processes and the processes which include multiple soft scattering. 

\begin{figure}[!h]
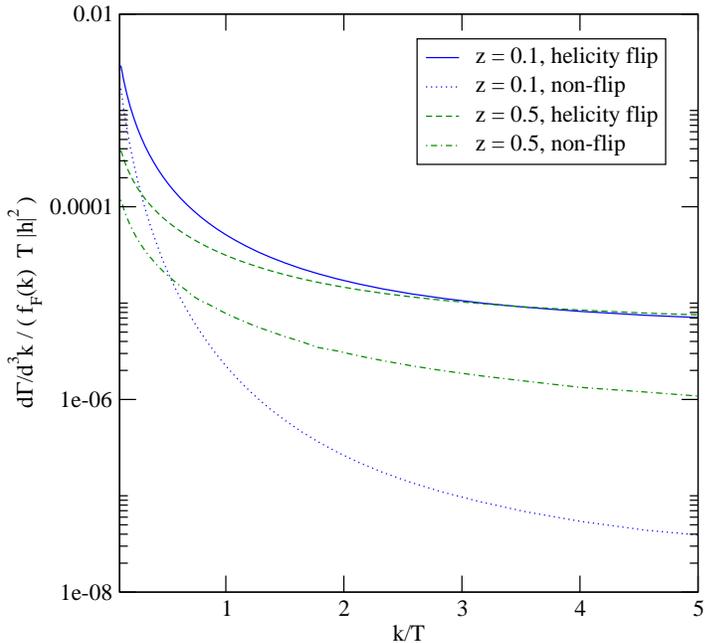

 \centering
 \centerbox{0.6}{dCollineardkHeli_new.eps} \vspace{0.5cm}
 \caption{Differential production rate divided by Fermi-Dirac distribution 
function for two different temperatures.}
 \label{fg:spectrum}
\end{figure}

Finally  we study the momentum spectrum of the produced heavy neutrinos.
If it was thermal, the rate would be proportional to the Fermi-Dirac
distribution $ f _ { \rm F } ( k ) $.  In Fig.~\ref{fg:spectrum} we
show the ratio of the differential production rate and $ f _ { \rm F }
( k ) $ for two different temperatures. For $z = 0.1$ the production
occurs both via a tree-level process (in this case, the decay of the
Higgs boson) and via processes involving multiple soft scattering whereas for $z
= 0.5$ the tree-level processes are kinematically forbidden.
We see that the spectrum is peaked in the infrared which means that the
typical momenta of the produced neutrinos are smaller than in
equilibrium. This holds both for helicity changing  
and for helicity conserving processes. 

\section{Summary and Conclusions}
\label{sc:conclusions}

In this paper we have studied the production of relativistic Majorana
neutrinos, relevant for models of thermal leptogenesis, in a hot electroweak
plasma. Based on our previous work~\cite{BesakBodeker}, we have obtained an
equation which sums all leading order 
collinear production processes. It includes the
tree-level processes--the decay of the Higgs boson as well as the inverse decay of
the Majorana neutrino--and in addition 
processes involving multiple scattering mediated by soft gauge
boson exchange. All these turn out to contribute at leading order in the
coupling constants and have never been included in previous treatments of
thermal leptogenesis. 

Numerically, we find a very pronounced increase in the
thermal production rate when soft gauge interactions are included. 
 At high temperatures, when the production via tree-level Higgs decay is
 allowed, the
rate increases by about a factor 3. When the tree-level 
processes are forbidden, the rate in units of temperature drops only slightly,
and the production of Majorana neutrinos remains effective.
When the temperature is close to  $M_N$ the rate is not significantly affected
by soft gauge interactions. We further showed that the 
production rate is dominated by helicity changing processes and that helicity 
conserving
(inverse) decays only play a significant role at low temperatures, $T \sim M_N$.
In addition to the production rate, we have studied the momentum spectrum of 
the produced 
Majorana neutrinos and found that it is strongly peaked in the infrared. 
The spectrum is thus not even approximately thermal. 

The leading order production rate of heavy Majorana neutrinos also receives
contributions from $ 2 \leftrightarrow 2 $ scattering processes. They have
been considered previously in the context of thermal leptogenesis
(see e.g.\ \cite{Giudice,pilaftsis,pedestrians,davidson,hahn-woernle}), but a
complete and consistent leading order
calculation has not been done so far.  In \cite{Giudice} the Higgs decay was
found to do\-minate over the $ 2 \leftrightarrow 2 $ processes at high
temperature. In this paper we have shown that adding soft gauge interactions
to the Higgs decay process leads to a strong enhancement of the production
rate. Therefore it would be very interesting to see how the processes
discussed here, together with the $ 2 \leftrightarrow 2 $ scattering processes
affect various scenarios of leptogenesis.

\vspace{.5cm}
{\bf Acknowledgments} 
This work was supported in part through the
DFG funded Graduate School GRK 881.

\appendix 
\renewcommand{\theequation}{\thesection.\arabic{equation}}
\setcounter{equation}0

\section{Solving the integral equations}   
\label{app:Diffeq_LPM}

The $ \vec p _\perp $-integrals of the functions $ \vec f $ and $ \psi  $ 
in Eq.~(\ref{ratepsif}) can be obtained without solving the integral equations 
\eqref{inteq_forf} 
and \eqref{inteq_forpsi}. This is achieved by Fourier transformation which 
turns Eq.~\eqref{inteq_forf} 
and \eqref{inteq_forpsi} into boundary value problems. Our approach which 
is described in this Appendix closely follows Refs.~\cite{Aurenche1,AurencheDilepton}. 

\subsection{From integral equation to differential equation}

Eqs.~\eqref{inteq_forf} and \eqref{inteq_forpsi} are  
of the same form as the integral equation for the production of transverse 
and of longitudinal photons.  
Following~\cite{Aurenche1,AurencheDilepton} we   Fourier
transform \footnote{We use the same symbol for  the functions $\vec
  f,\psi$ and their Fourier transforms.} 
\begin{align}\label{fouriertrafo_of_psi}
 \vec f(\vec b ) = \int \text{d}^2 b \, e^{i\vec p_\perp \cdot \vec b}
  \vec f (\vec p _\perp )
  , \qquad 
  \psi(\vec b) = \int \text{d}^2 b\, e^{i\vec p_\perp \cdot \vec b} \psi(\vec p
  _\perp ).
\end{align}
Then the integral over the perpendicular components in
\eqref{ratepsif} can be replaced by the limit of their Fourier
transforms for $b \to 0$: 
\begin{align}
 \label{fptofb} \int \frac{\text{d}^2 p_\perp}{(2\pi)^2} \operatorname{Re} [\vec p_\perp \cdot
  \vec f(\vec p_\perp)] &= \lim_{b \to 0} \operatorname{Im} \vec \nabla \cdot \vec f(\vec b)
  , \\
\int \frac{\text{d}^2 p_\perp}{(2\pi)^2} \operatorname{Re} \psi(\vec p_\perp)
&= \lim_{b \to 0} \operatorname{Re} \psi(\vec b) \hfill 
\end{align}  
The integral equations (\ref{inteq_forf}), (\ref{inteq_forpsi}) now turn into
differential equations for 
$ \vec f ( \vec b ) $ and $\psi(\vec b)$. 
With $ \vec k _\perp = \vec 0 $ and $ k ^ 2 = M _ N ^ 2 $ we can write  
\begin{align}
  \epsilon(k,\vec p) = \beta \left( \vec p_\perp ^2 + M_{\rm eff} ^2 \right)
\end{align}
where
\begin{align}
  \beta = \frac{k_\parallel}{2p_\parallel (p_\parallel - k_\parallel)}, 
  \quad M_{\rm eff} ^ 2 = \frac{p_\parallel (p_\parallel - k_\parallel)M_N ^2 
    - k_\parallel(p_\parallel - k_\parallel)m_\ell ^2 + k_\parallel p_\parallel
    m_\varphi ^2 }
  {k_\parallel ^2}
1  .
\end{align}
Then we obtain differential equations which closely resemble those in \cite{AurencheDilepton},
\begin{align}
  \label{diffeq2} 
  -i\beta \left( \Delta - M_{\rm eff} ^2
    \right) \vec f(\vec b) -  \mathcal{K}(b) \vec f(\vec b) 
    =& \, -i 2 \vec \nabla \delta(\vec b)
    , \hfill 
  \\
  \label{diffeq1} 
  -i\beta \left( \Delta -
    M_{\rm eff} ^2 \right)\psi(\vec b)  - \mathcal{K}(b) \psi(\vec b) =& \, \delta(\vec b) \hfill 
\end{align}
where $ b \equiv | \vec b | $, and 
\be 
\mathcal{K}(b) = T\left[ C_2(r)g^2 D(m_{\rm D} b) + y_\ell ^2
    {g'} ^2 D(m_{\rm D} ' b)\right]
\ee
with~\cite{Aurenche1,AurencheDilepton}
\be
D(x) = \frac{1}{2\pi} \left[ \gamma_{\rm E} + \ln \frac{x}{2} + K_0(x) \right].
\label{functionD}
\ee
Here, $\gamma_{\rm E} = 0.5772...$ denotes the Euler-Mascheroni constant and
$K_0(x)$ is a modified Bessel function. 

The convergence of the Fourier integrals enforces the boundary conditions
\begin{align}\label{bcatinfinity}
  \lim_{b \to \infty} \vec f(\vec b) = 0, \quad 
  \lim_{b \to \infty} \psi(\vec b) = 0
  .
\end{align}
In the limit $b \to 0$ all terms
which contain no derivatives can be neglected in (\ref{diffeq2}), 
(\ref{diffeq1}),  and we obtain the limiting
behaviour 
\begin{align}
  \label{fsmall} \vec f(\vec b) =& \, 
  c _ {  f } 
  \frac{\vec b}{b^2} +
  O( b)
  ,
  \\ 
  \label{psismall} \psi(\vec b) =& \, 
  c _ \psi  \ln b +
  O(b^0) \hfill 
  . \hfill  
\end{align}
The constants $ c _ f $, $ c _ \psi  $
are fixed by the $ \delta $-functions in Eqs.~(\ref{diffeq2}),
(\ref{diffeq1}) which gives  
\begin{align}
  \label{f_small} c _ f  =& \, \frac{1}{\pi \beta} 
  \, ,
  \\ 
  \label{psi_small} c _ \psi =& \, \frac{i}{2\pi \beta} 
  \, . \hfill  
\end{align}

Due to  rotational invariance we must have $\vec f(\vec b) = h(b)\vec b$ 
 and $ \psi ( \vec b ) = \psi ( b ) $. In terms of the new function $h(b)$
the RHS of \eqref{fptofb} becomes $2\lim_{b \to 0} \operatorname{Im} h(b)$.
We numerically solve the  ODEs 
for $ h ( b ) $ and $ \psi  ( b ) $ 
\begin{align}
  \label{diffeq_forh} 
  -i\beta \left( \partial_b ^2 + \frac{3}{b} \partial_b -
    M_{\rm eff} ^2 \right) h(b) - \mathcal{K}(b) h(b) =& \, 0 
  \hfill 
  \hfill \\ 
  \label{diffeq_forpsi} 
  -i\beta \left( \partial_b ^2 + \frac{1}{b} \partial_b -
    M_{\rm eff} ^2 \right) \psi(b) - \mathcal{K}(b) \psi(b) =& \, 0 
\end{align}
for $b > 0$.
The conditions \eqref{bcatinfinity}, \eqref{f_small} and \eqref{psi_small} 
are then sufficient to determine the solutions unambiguously.


\subsection{Numerical procedure} 

Finally we want to describe our algorithm that we used to obtain the solutions
shown in Sec.~\ref{sc:Numerical}. 
Both equations \eqref{diffeq_forpsi} and \eqref{diffeq_forh} can be solved by
the same method and we only describe the procedure in terms of \eqref{diffeq_forpsi}
explicitly.

For the numerical solution, it proves convenient to split the function 
into a tree-level part and another part coming from multiple soft scattering,
\be 
\psi(b) = \psi_0(b) + \psi_1(b).
\ee
The function $\psi_0$ solves \eqref{diffeq1} with $D(b) \equiv 0$ and can be written 
in terms of Bessel functions. 
For $b \to 0$, the function $D(b)$ behaves like $D(b) \sim b^2 \ln b$. This implies
that the  limiting behaviour of $ \psi  _ 0 $ for $b \to 0$ is also given by
Eq.~\eqref{psi_small}. Therefore $\psi_1(b)$ 
must be regular at $b = 0$. The general solution for $ \psi  _ 1 $ can be written as
\be \label{psi_general}
\psi_1 (b) = c_1 \psi_1 ^{(1)}(b) + c_2 \psi_1 ^{(2)} (b) + \psi_1 ^{\rm (part)}(b)
\ee
where $\psi_1 ^{\rm (part)}$ is a particular solution of the inhomogeneous
equation while the linearly independent solutions
$\psi_1 ^{(i)}$ solve the corresponding homogeneous
equation. One can choose $ \psi_1  ^{ ( 1 ) } $ such that it has the limiting
behavior of Eq.~\eqref{psismall}, and a  $ \psi_1  ^{ ( 2 ) } $ which is regular for $
b \to 0 $. Thus we must have $ c _ 1 = 0 $. 
The other
constant is then found by imposing \eqref{bcatinfinity} and we obtain
\be \label{c2}
c_2 = -\lim_{b \to \infty} \frac{\psi_1 ^{\rm (part)}(b)}{\psi_1 ^{(2)} (b)}.
\ee
Numerically, one has to choose finite values for '$b \to 0$' and '$b \to
\infty$'. We have found $b_0 = 10^{-5}T^{-1}$ and $b_\infty = 30T^{-1}$ to be a
reasonable choice, giving stable numerical solutions.

The algorithm providing us with the desired solution is then the following:
\begin{enumerate}
 \item Solve the \emph{homogeneous} ODE for $\psi_1(b)$ with $\psi_1(b_0) = 1,
   \psi_1 '(b_0) = 0$ to obtain $\psi_1 ^{(2)} (b)$, 
 \item Solve the \emph{inhomogeneous} ODE for $\psi_1(b)$ with  $\psi_1(b_0) =
   i, \psi_1 '(b_0) = 0$ to obtain $\psi_1 ^{\rm (part)} (b)$, 
 \item Compute $c_2$ via \eqref{c2}. Then, due to the 
initial conditions chosen in the  previous steps, the  desired result is given by
   $\operatorname{Re} \psi_1(b \to 0) = \operatorname{Re} c_2 $. 
\end{enumerate}
In order to find $\operatorname{Im} h_1(b \to 0)$ we can use the same
algorithm, only in the second step it is
more convenient to choose $h_1(b_0) = 1 $, $ h_1'(b_0) = 0$ because then 
$\operatorname{Im} h_1(b \to 0) = \operatorname{Im} c_2$.


\end{document}